\documentclass[oneside,a4paper]{article}

\usepackage{setspace}
\usepackage{amsmath,amssymb}
\usepackage[T1]{fontenc}
\usepackage[polish,english]{babel}
\usepackage[utf8]{inputenc}
\usepackage{url}
\usepackage{calc}
\usepackage{graphicx}
\usepackage{fancyhdr}
\usepackage{color}
\usepackage{natbib}
\usepackage{multirow}
\usepackage{comment}
\usepackage{array}
\usepackage{tabulary,tabularx}
\usepackage{threeparttable}
\usepackage{lscape}
\usepackage{ragged2e}
\usepackage{float}
\restylefloat{table}

\newcolumntype{P}[1]{>{\RaggedRight\hspace{0pt}}p{#1}}
\newcolumntype{L}[1]{>{\raggedright\let\newline\\\arraybackslash\hspace{0pt}}m{#1}}
\newcolumntype{C}[1]{>{\centering\let\newline\\\arraybackslash\hspace{0pt}}m{#1}}
\newcolumntype{R}[1]{>{\raggedleft\let\newline\\\arraybackslash\hspace{0pt}}m{#1}}

\newcommand{\ud}{\mathrm{d}}

\begin{document}

\title{\large \textbf{Bayesian analysis of seasonally cointegrated VAR model}}

\date{}

\author{Justyna Wróblewska}

\author{Justyna Wróblewska\thanks{Cracow University of Economics, Department of Econometrics and Operational Research, Rakowicka 27, 31-510 Kraków, Poland, e-mail: eowroble@cyf-kr.edu.pl}}

\maketitle
\thispagestyle{fancy}
\fancyhead{}

\begin{abstract}
The paper aims at developing the Bayesian seasonally cointegrated model for quarterly data. We propose the prior structure, derive the set of full conditional posterior distributions, and propose the sampling scheme. The identification of cointegrating spaces is obtained \emph{via} orthonormality restrictions imposed on vectors spanning them. In the case of annual frequency, the cointegrating vectors are complex, which should be taken into account when identifying them. The point estimation of the cointegrating spaces is also discussed. The presented methods are illustrated by a simulation experiment and are employed in the analysis of money and prices in the Polish economy.

\end{abstract}

\begin{center} \end{center}
\textbf{Keywords:} seasonal cointegration, reduced rank regression, error correction model, Bayesian analysis, Bayesian model comparison;

\begin{center} \end{center}
\textbf{JEL Classification:} C11, C32, C53;

%\begin{center} \end{center}
%\textbf{Acknowledgements:}
%\\.

\newpage

\section{Introduction}
\label{sec:intro}

Many macroeconomic quarterly or monthly time series display strong both trend and seasonal behavior. The idea of cointegration at zero frequency (assuming common stochastic trend controlling the long-run behavior of the series) is well known and oft employed in the empirical analyses. Methods enabling estimation of parameters of vector error correction models are well known within both classical (see e.g. \citealp{Johansen1995}) and Bayesian paradigm (see e.g. \citealp{Koop_al2006}). However, although the idea of cointegration at seasonal frequencies was introduced in 1990 (\citealp{Hylleberg_al1990}) it got much fewer interests. In most researches, the seasonally adjusted data are analyzed or the seasonality is modeled \emph{via} seasonal dummies. However, there are papers presenting the unwilling results of seasonal adjustment. Such procedures may for example change both the short- and long-run behavior of the series (e.g. \citealp{Nerlove1964}, \citealp{Cubadda1999}, \citealp{Hecq1998}, \citealp{GrangerSiklos1995}).\footnote{Note that the very same problems appear in the case of filtering non-seasonal unit roots (see e.g. \citealp{MeyerWinker2005}, \citealp{Hamilton2018}).}
\cite{Abeysinghe1994} discusses the problems which may occur in researchers where seasonality is modeled by seasonal dummies in cases when such behavior is generated by seasonal unit roots.  Moreover, \cite{LofFranses2001} shows that taking into account seasonal cointegration improves forecast abilities of models.\\
The likelihood method enabling estimation of seasonal VEC models was proposed by \cite{Johansen_Schaumburg1999} (see also \citealp{Cubadda_Omtzigt2005} for further discussion), but to the best of our knowledge, there are no Bayesian seasonal VEC models. This paper is aimed to fulfill this gap by building a Bayesian model for the quarterly seasonally cointegrated time series. Priors for the parameters are imposed. The set of full conditional posterior distributions is obtained. Additionally, the point estimation of the cointegrating spaces is discussed. The proposed methods are illustrated by a small simulation experiment and the empirical analysis of money and prices in the Polish economy conducted in the four-dimensional seasonal system.

\section{Bayesian seasonal VEC model}
\label{sec:model}

We start the analysis with the assumption that the $n-$dimensional quarterly time series $\{y_t\}$ has the following VAR($k$) representation:
\begin{equation}
y_t=\sum_{j=1}^kA_jy_{t-j}+\Phi D_t+\varepsilon_t,\quad \varepsilon_t\sim iiN(0,\Sigma)
\label{eqn:VAR}
\end{equation}
where $D_t$ contains deterministic components such as a constant, a trend or seasonal dummies. The initial conditions - $y_0, y_{-1}, \dots, y_{-k+1}$ - are fixed, $A(L)=I_n-A_1L-\dots-A_kL^k$ is the polynomial matrix of the process (\ref{eqn:VAR}).\\
According to Lagrange expansion the polynomial $A(z)$ around the points $z_1,z_2,\dots,z_S$ may be written as (\citealp{Johansen_Schaumburg1999}, \citealp{Kotlowski2005}):
\begin{equation}
A(z)=p(z)I_n+\sum_{s=1}^SA(z_s)\frac{p_s(z)z}{p_s(z_s)z_s}+p(z)zA_0(z),
\label{eqn:Lagrange}
\end{equation}
where $p(z)=\prod_{s=1}^S(1-\bar{z}_sz)$, $p_j(z)=\prod_{s\neq j}^S(1-\bar{z}_sz)=\frac{p(z)}{1-\bar{z}_jz},\ z\neq z_j$ and $A_0(z)$ is a polynomial matrix. If additionally $z_s$ is a root of the characteristic polynomial of the process (\ref{eqn:VAR}), i.e. $|A(z_s)|=0$ then $A(z_s)$ is of the reduced rank and can be decomposed as the product of full column rank matrices $A(z_s)=a_sb_s'$.\\
The expansion (\ref{eqn:Lagrange}) leads for processes cointegrated at zero and quarterly frequencies (i.e. $|A(z)|=0$ for $z_1=1$, $z_2=-1$, $z_3=i$, $z_4=\bar{z}_3=-i$) to the following seasonal error correction representation (see e.g. \citealp{Hylleberg_al1990}, \citealp{Johansen_Schaumburg1999}, \citealp{Cubadda_Omtzigt2005}, \citealp{Kotlowski2005}):
\begin{eqnarray}
\Delta_4y_t&=&\alpha_1\beta_1'\tilde{y}_t^{(1)}+\alpha_2\beta_2'\tilde{y}_t^{(2)}+\alpha_{\star}\bar{\beta}_{\star}'\tilde{y}_t^{(3)}+\bar{\alpha}_{\star}\beta_{\star}'\bar{\tilde{y}}_t^{(3)}+\nonumber\\
&&{}+\sum_{i=1}^{k-4}\Gamma_j\Delta_4y_{t-j}+\tilde{\Phi}\tilde{D}_t+\varepsilon_t,\quad \varepsilon_t\sim iiN(0,\Sigma),
\label{eqn:basicC}
\end{eqnarray}
where $\Delta_4y_t=(1-L^4)y_t=y_t-y_{t-4}$ with $L$ denoting the lag operator, $\Gamma_j=-\sum_{l=1}^{[(k-j)/4]}A_{j+4l},\ j=1,2,\dots,k-4$. Vectors $\tilde{y}_t^{(\cdot)}$ may contain deterministic components, so $\tilde{y}_t^{(\cdot)}=\left(\begin{matrix}y_t^{(\cdot)}\\d_t^{(\cdot)}\end{matrix}\right)$. In particular, let consider the deterministic component of the following form $\Phi D_t=\mu+a\cos\left(\frac{\pi}{2}t\right)+b\sin\left(\frac{\pi}{2}t\right)+c\cos\left(\pi t\right)+\gamma t$ (see \citealp{Franses_Kunst1999} and \citealp{Kotlowski2005}).\\
Employing the polynomials $p_{\cdot}(L)$ to the  stochastic and deterministic parts of the process (\ref{eqn:VAR}) leads to the vectors $\tilde{y}_t^{(\cdot)}$ of the following forms:
\begin{itemize}
\item at zero frequency -  $\tilde{y}_t^{(1)}=\left(\begin{matrix}y_t^{(1)}\\d_t^{(1)}\end{matrix}\right)$, where $y_t^{(1)}=p_1(L)Ly_t=(1+L+L^2+L^3)Ly_t=y_{t-1}+y_{t-2}+y_{t-3}+y_{t-4}$, the term $(1+L+L^2+L^3)LD_t$ leads to $d_t^{(1)}=t-\frac{5}{2}$ and an unrestricted constant,
\item at $\pi$ frequency -  $\tilde{y}_t^{(2)}=\left(\begin{matrix}y_t^{(2)}\\d_t^{(2)}\end{matrix}\right)$, where $y_t^{(2)}=p_2(L)Ly_t=(1-L+L^2-L^3)Ly_t=y_{t-1}-y_{t-2}+y_{t-3}-y_{t-4}$, the term $(1-L+L^2-L^3)LD_t$ leads to $d_t^{(2)}=cos(\pi t)$ and an unrestricted constant,
\item at $\frac{\pi}{2}$ and $\frac{3\pi}{2}$ frequencies -  $\tilde{y}_t^{(3)}=\left(\begin{matrix}y_t^{(3)}\\d_t^{(3)}\end{matrix}\right)$, where $y_t^{(3)}=p_3(L)Ly_t=(-i-L+iL^2+L^3)Ly_t=-iy_{t-1}-y_{t-2}+iy_{t-3}+y_{t-4}$,  the term $(-i-L+iL^2+L^3)LD_t$ leads to $d_t^{(3)}=\cos\left(\frac{\pi}{2}t\right)-i\sin\left(\frac{\pi}{2}t\right)$ and an unrestricted constant.
\end{itemize}
Note that the unrestricted constant occurring in each of the above considered frequencies results form the linear trend assumed for the level of the analyzed process (\ref{eqn:VAR}). If there is no linear trend, but only constant, there is neither trend restricted to the cointegration space at the zero frequency nor an unrestricted constant in the representation (\ref{eqn:basicC}), but there is a constant restricted to the cointegration spaces of the zero frequency ($d_t^{(1)}=1$), see e.g. \citet[pp. 93-112]{Juselius2006} for the discussion of the meaning of dummies gathered in the vectors $D_t$ and $\tilde{D}_t$, and also the relations between them.\\
Note also that $\alpha_{\star}\bar{\beta}_{\star}'\tilde{y}_t^{(3)}$ and $\bar{\alpha}_{\star}\beta_{\star}'\bar{\tilde{y}}_t^{(3)}$ are complex conjugate matrices, so their sum gives their real part multiplied by 2: 
\begin{eqnarray}
\alpha_{\star}\bar{\beta}_{\star}'\tilde{y}_t^{(3)}+\bar{\alpha}_{\star}\beta_{\star}'\bar{\tilde{y}}_t^{(3)}&=&2 Re(\alpha_{\star}\bar{\beta}_{\star}'\tilde{y}_t^{(3)})=\\
&=&2\left[(\alpha_I\beta_R'-\alpha_R\beta_I')(y_{t-1}-y_{t-3})-(\alpha_R\beta_R'+\alpha_I\beta_I')(y_{t-2}-y_{t-4})\right],\nonumber
\label{eq:conj}
\end{eqnarray}
where $\alpha_R$, $\beta_R$ denote the real parts of $\alpha_{\star}$ and $\beta_{\star}$ respectively, whereas $\alpha_I$, $\beta_I$ - their imaginary parts ($\alpha_{\star}=\alpha_R+i\alpha_I$, $\beta_{\star}=\beta_R+i\beta_I$).\\
These leads to the more commonly used representation of seasonally cointegrated quarterly VAR process (see e.g. \citealp{Hylleberg_al1990}, \citealp{Johansen_Schaumburg1999}, \citealp{Cubadda_Omtzigt2005}, \citealp{Kotlowski2005}):
\begin{equation}
\Delta_4y_t=\Pi_1\tilde{y}_t^{(1)}+\Pi_2\tilde{y}_t^{(2)}+\Pi_3\tilde{y}_t^{(32)}+\Pi_4\tilde{y}_t^{(31)}+\sum_{i=1}^{k-4}\Gamma_i\Delta_4y_{t-i}+\tilde{\Phi}\tilde{D}_t+\varepsilon_t,\quad \varepsilon_t\sim iiN(0,\Sigma),
\label{eqn:basic}
\end{equation}
where $\Pi_1=\alpha_1\beta_1'$, $\Pi_2=\alpha_2\beta_2'$, $\Pi_3=-2(\alpha_R\beta_R'+\alpha_I\beta_I')$, $\Pi_4=2(\alpha_I\beta_R'-\alpha_R\beta_I')$, $\tilde{y}_t^{(31)}=\left(\begin{matrix}y_t^{(31)}\\d_t^{(31)}\end{matrix}\right)$, where $y_t^{(31)}=(1-L^2)Ly_t=y_{t-1}-y_{t-3}$, $d_t^{(31)}=sin(\frac{\pi t}{2})$, $\tilde{y}_t^{(32)}=\left(\begin{matrix}y_t^{(32)}\\d_t^{(32)}\end{matrix}\right)$, where $y_t^{(32)}=(1-L^2)L^2y_t=y_{t-2}-y_{t-3}$, $d_t^{(32)}=cos(\frac{\pi t}{2})$.

To save on notation we introduce the matrix form of the model (\ref{eqn:basicC}).
\begin{equation}
Z_0=Z_1\beta_1\alpha_1'+Z_2\beta_2\alpha_2'+Z_3\bar{\beta}_{\star}\alpha_{\star}'+\bar{Z}_3\beta_{\star}\bar{\alpha}_{\star}'+Z_4\Gamma+E,
\label{eqn:basicC_matrix}
\end{equation}
where $Z_0=\left(\begin{matrix}\Delta_4y_1&\Delta_4y_2&\dots&\Delta_4y_T\end{matrix}\right)'$, $Z_1=\left(\begin{matrix}\tilde{y}_1^{(1)}&\tilde{y}_2^{(1)}&\dots&\tilde{y}_T^{(1)}\end{matrix}\right)'$, $Z_2=\left(\begin{matrix}\tilde{y}_1^{(2)}&\tilde{y}_2^{(2)}&\dots&\tilde{y}_T^{(2)}\end{matrix}\right)'$, $Z_3=\left(\begin{matrix}\tilde{y}_1^{(3)}&\tilde{y}_2^{(3)}&\dots&\tilde{y}_T^{(3)}\end{matrix}\right)'=-Z_{32}-iZ_{31}$, $Z_{31}=\left(\begin{matrix}\tilde{y}_1^{(31)}&\tilde{y}_2^{(31)}&\dots&\tilde{y}_T^{(31)}\end{matrix}\right)'$, $Z_{32}=\left(\begin{matrix}\tilde{y}_1^{(32)}&\tilde{y}_2^{(32)}&\dots&\tilde{y}_T^{(32)}\end{matrix}\right)'$,  $Z_4=\left(\begin{matrix}z_1&z_2&\dots&z_T\end{matrix}\right)'$, $z_t'=\left(\begin{matrix}\Delta_4y_{t-1}'&\Delta_4y_{t-2}'&\dots&\Delta_4y_{t-k+4}'&\tilde{D}_t\end{matrix}\right)$, $\Gamma=\left(\begin{matrix}\Gamma_1&\Gamma_2&\dots&\Gamma_{k-4}&\tilde{\Phi}\end{matrix}\right)'$, $E=\left(\begin{matrix}\varepsilon_1&\varepsilon_2&\dots&\varepsilon_T\end{matrix}\right)'$.\\
As the analyzed data inform only about the cointegration space not the cointegration vectors, during the estimation we have to deal with the non-identification occurring in the products: $\alpha_1\beta_1'$, $\alpha_2\beta_2'$, $\bar{\beta}_{\star}\alpha_{\star}'$, $\beta_{\star}\bar{\alpha}_{\star}'$. Therefore, we employ the methods proposed by \citet{Koop_al2009}. Following their ideas we will consider two observationally equivalent representation for each of the considered products. In the $A-B$ representations it is assumed that the matrices belong to the $\mathbb{R}^{\cdot}$ or $\mathbb{C}^{\cdot}$ spaces of appropriate dimensions, whereas in the $\alpha-\beta$ representations $\beta$s have orthonormal columns and $\alpha$s still belong tho the $\mathbb{R}^{\cdot}$ or $\mathbb{C}^{\cdot}$ spaces:
\begin{itemize}
\item $A_1B_1'\equiv\alpha_1\beta_1',$\\$A_1\in\mathbb{R}^{n\times r_1}, B_1\in\mathbb{R}^{m_1\times r_1}, \alpha_1=A_1(B_1'B_1)^{\frac{1}{2}}\in\mathbb{R}^{n\times r_1}, \beta_1=B_1(B_1'B_1)^{-\frac{1}{2}}\in\mathbb{V}_{r_1,m_1}$,
\item $A_2B_2'\equiv\alpha_2\beta_2',$\\$A_2\in\mathbb{R}^{n\times r_2}, B_2\in\mathbb{R}^{m_2\times r_2}, \alpha_2=A_2(B_2'B_2)^{\frac{1}{2}}\in\mathbb{R}^{n\times r_2}, \beta_2=B_2(B_2'B_2)^{-\frac{1}{2}}\in\mathbb{V}_{r_2,m_2}$,
\item $A_{\star}\bar{B}_{\star}'\equiv\alpha_{\star}\bar{\beta}_{\star}',$\\$ A_{\star}=A_R+iA_I\in\mathbb{C}^{n\times r_3}, B_{\star}=B_R+iB_I\in\mathbb{C}^{m_3\times r_3}, \alpha_{\star}=A_{\star}(\bar{B}_{\star}'B_{\star})^{\frac{1}{2}}\in\mathbb{C}^{n\times r_3}, \beta_{\star}=B_{\star}(\bar{B}_{\star}'B_{\star})^{-\frac{1}{2}}\in\mathbb{V^C}_{r_3,m_3}$,
\end{itemize}  
where $\mathbb{V}_{r_j,m_j},\ j=1,2$ denotes the Stiefel manifold, i.e. the set of $m_j\times r_j$ matrices with orthonormal columns ($\mathbb{V}_{r_j,m_j}=\left\{X\in\mathbb{R}^{m_j\times r_j}: X'X=I_{r_j}\right\}$ ), $\mathbb{V^C}_{r_3,m_3}$ stands for the complex Stiefel manifold, i.e. the set of $m_3\times r_3$ semi-unitary matrices ($\mathbb{V^C}_{r_3,m_3}=\left\{X\in\mathbb{C}^{m_3\times r_3}: \bar{X}'X=I_{r_3}\right\}$).\\
Note that by this approach the non-identification issue is only partially solved as there is many-to-one relationship between the Stiefel manifolds and the Grassmann manifolds \footnote{$\mathbb{G}_{r_j,m_j-r_j}, j=1,2,\ \mathbb{G^C}_{r_3,m_3-r_3}$, collecting $r_j$-dimensional planes, passing through the origin, in the real ($j=1,2$)/complex ($j=3$) vector $m_j$-dimensional space, see e.g. \citealp{James1954}, \citealp{Chern_Wolfson1987}} to which belong the cointegration spaces: if $X$ is the element of the (complex) Stiefel manifold and the $r_j\times r_j$ ($j=1,2,3$) matrix $O$ is the element of the (complex) orthonormal group ($O'O=OO'=I_{r_j},\ j=1,2$, $\bar{O}'O=O\bar{O}'=I_{r_3}$) than $XO$ is the element of the same (complex) Stiefel manifold and they span the same spaces (the projection matrices are equal, i.e. $XX'=XOO'X'$ in the real case and $X\bar{X}'=XO\bar{O}'\bar{X}'$ in the complex case).\\
The first two products (i.e. $\alpha_1\beta_1$ and $\alpha_2\beta_2$) involve only matrices with real numbers, so can be treated exactly as proposed by \citet{Koop_al2009}. While in the third case we have to adjust their procedures to the complex matrices and spaces.\\
We start the analysis with the $A-B$ parameterization and impose the following prior distributions over the model parameters:
\begin{itemize}
\item the inverse Wishart distribution fo the covariance matrix - $\Sigma\sim iW(S,q),$

\item the matrix normal distribution for $\Gamma$ - $\Gamma|\Sigma,\nu\sim mN(\underline{\mu}_{\Gamma},\Sigma,\nu\underline{\Omega}_{\Gamma}),$

\item the matrix normal distribution for adjustment coefficients at frequency 0 - $A_1|\Sigma,\nu\sim mN(\underline{\mu}_1,\nu\underline{\Omega}_1,\Sigma),$ 

\item the matrix normal distribution for un-normalized cointegrating vectors at frequency 0 - $B_1\sim mN(0,\frac{1}{m_1}I_{r_1},P_1)$, which leads to matrix angular central distribution for its orientation - $\beta_1\sim MACG(P_1)$ (see \citealp{Chikuse1990}, \citeyear{Chikuse2003}), \emph{via} the matrix $P_1$ the researcher can incorporate prior knowledge about the cointegration space at zero frequency (see \citealp{Koop_al2009} for the details),

\item the matrix normal distribution for adjustment coefficients at frequency $\pi$ - $A_2|\Sigma,\nu\sim mN(\underline{\mu}_2,\nu\underline{\Omega}_2,\Sigma),$

\item the matrix normal distribution for un-normalized cointegrating vectors at frequency $\pi$ - $B_2\sim mN(0,\frac{1}{m_2}I_{r_2},P_2)$, so $\beta_2\sim MACG(P_2)$ (see the explanation stated in the point for $B_1$),

\item the complex matrix normal distribution for adjustment coefficients at frequencies $\frac{\pi}{2}$ and $\frac{3\pi}{2}$ - $A_{\star}|\Sigma,\nu\sim mCN(\underline{\mu}_{\star},\nu I_{r_3},\Sigma),$ i.e. $p(A_{\star}|\Sigma)=\pi^{-nr_3}|\Sigma|^{-r_3}\exp\{-tr\Sigma^{-1}(A_{\star}-\underline{\mu}_{\star})(\frac{1}{\nu}I_{r_3})({\bar{A}_{\star}-\bar{\underline{\mu}}}_{\star})'\}$, so $E(A_{\star})=\underline{\mu}_{\star}$, $V(vec(A_{\star}))=I_{r_3}\otimes\Sigma$, where $\underline{\mu}_{\star}=\underline{\mu}_{\star R}+i\underline{\mu}_{\star I}$. Note that imposing such distribution for $A_{\star}$ is equivalent to assuming that $\left(\begin{matrix}A_R\\A_I\end{matrix}\right)|\Sigma\sim mN\left(\left(\begin{matrix}\underline{\mu}_{\star R}\\\underline{\mu}_{\star I}\end{matrix}\right),\nu I_{r_3},\left(\begin{matrix}\frac{1}{2}\Sigma&\mathbf{0}_{n\times n}\\ \mathbf{0}_{n\times n}&\frac{1}{2}\Sigma\end{matrix}\right)\right)$,

\item the complex matrix normal distribution for un-normalized cointegrating vectors at frequencies $\frac{\pi}{2}$ and $\frac{3\pi}{2}$ - $B_{\star}\sim mCN(\mathbf{0}_{m_r\times r_3},\frac{1}{m_3}I_{r_3},P_{\star})$, where $P_{\star}=P_{\star R}+iP_{\star I}$, such as $P_{\star}$ is Hermitian ($P_{\star}=\bar{P}_{\star}'\equiv (P_{\star R}=P_{\star R}', P_{\star I}=-P_{\star I}')$) positive definite matrix, so for the real and imaginary part of $B_{\star}$ we impose the matrix normal distribution of the following form $\left(\begin{matrix}B_R\\B_I\end{matrix}\right)\sim mN\left(0_{2m_3\times r_3},I_{r_3},\frac{1}{2}\left(\begin{matrix}P_{\star R}&-P_{\star I}\\P_{\star I}&P_{\star R}\end{matrix}\right)\right)$. Such distribution leads to the complex matrix angular central Gaussian distribution for the orientation part of the matrix $B^{\star}$ (see \citealp{Wroblewska2020}).

\item  The parameter $\nu$ may be estimated or settled by the researcher. In the case of estimated $\nu$ we propose to impose the inverse gamma distribution for it - $\nu\sim iG(\underline{s}_{\nu},\underline{n}_{\nu})$.
\end{itemize}
The joint prior distribution is truncated by the non-explosive condition taking into account the appropriate numbers of unit roots at each frequency.

The above proposed prior distributions together with the likelihood function lead to the joint posterior distribution with the following kernel:
\begin{eqnarray}
p(\theta|y)&\propto&\nu^{-\underline{n}_{\nu}-\frac{n}{2}[n(k-4)+l+r_1+r_2+2r_3]-1}\exp\left(-\frac{\underline{s}_{\nu}}{\nu}\right)|\Sigma|^{-[q+n(k-4)+l+r_1+r_2+2r_3+T+n+1]/2}\times\nonumber\\
&\times&\exp\left\{-\frac{1}{2}tr\left[\Sigma^{-1}\left[S+\frac{1}{\nu}\left(\Gamma-\underline{\mu}_{\Gamma}\right)'\underline{\Omega}_{\Gamma}^{-1}(\Gamma-\underline{\mu}_{\Gamma})+2\frac{1}{\nu}(A_{\star}-\underline{\mu}_{\star})(A_{\star}-\underline{\mu}_{\star})'\right]\right]\right\}\times\\
&\times&\exp\left\{-\frac{1}{2}tr\left[\Sigma^{-1}\left[\frac{1}{\nu}(A_1-\underline{\mu}_1)\underline{\Omega}_1^{-1}(A_1-\underline{\mu}_1)'+\frac{1}{\nu}(A_2-\underline{\mu}_2)\underline{\Omega}_2^{-1}(A_2-\underline{\mu}_2)'+E'E\right]\right]\right\}\times\nonumber\\
&\times&\exp\left\{-\frac{1}{2}tr\left[m_1B_1'P_1^{-1}B_1+m_3B_2'P_2^{-1}B_2+2m_3\bar{B}_{\star}'P_{\star}^{-1}B_{\star}\right]\right\}I_{[0,1]}(|\lambda|_{\max}).\nonumber
\label{eqn:post}
\end{eqnarray}
where $\theta=(\Sigma,\Gamma,A_1,A_2,A_{\star},B_1,B_2,B_{\star})$ collects all model's parameters, $l$ denotes the number of deterministic components gathered in $\tilde{D}_t$, $I_{[a,b]}(\cdot)$ is the indicator function of the interval $[a,b]$ and $\lambda$ stands for the vector of the eigenvalues of the companion matrix, that is the matrix of the form:
\begin{equation}
A=\left(\begin{matrix}
A_1&A_2&\ldots&A_{k-1}&A_k\\
I_n&0&\ldots&0&0\\
0&I_n&\ldots&0&0\\
\vdots&\vdots&\ddots&\vdots&\vdots\\
0&0&\ldots&I_n&0
\end{matrix}\right),
\end{equation}
where $I_n$ is an $n$-dimensional identity matrix, $A_1=\Pi_1+\Pi_2+\Pi_3+\Gamma_1$, $A_2=\Pi_1-\Pi_2+\Pi_4+\Gamma_2$, $A_3=\Pi_1+\Pi_2-\Pi_3+\Gamma_3$, $A_4=I_n+\Pi_1-\Pi_2-\Pi_4+\Gamma_4$, $A_i=\Gamma_i-\Gamma_{i-4}$ for $i=5,6,\dots$, $\Pi_1=\alpha_1\beta_1'\equiv A_1B_1'$, $\Pi_2=\alpha_2\beta_2'\equiv A_1B_1'$, $\Pi_3=2(\alpha_I\beta_R'-\alpha_R\beta_I')\equiv2(A_IB_R'-A_RB_I')$, $\Pi_4=-2(\alpha_R\beta_R'+\alpha_I\beta_I')\equiv-2(A_RB_R'+A_IB_I')$ and $\Gamma_i=0$ for $i>k-4$. The matrix A makes it possible to write the analyzed process in the VAR(1) form (see e.g. \citealp[pp. 15-16]{Lutkepohl2005}).\\
From equation (\ref{eqn:post}) we obtain the set of full conditional posterior distributions for model's parameters:
\begin{itemize}
\item the inverse Wishart distribution for the covariance matrix of errors $\Sigma$:
\begin{equation}
p(\Sigma|\cdot,y)=f_{iW}\left(\overline{S},q+n(k-4)+l+r_1+r_2+2r_3+T\right),
\label{eq:iW}
\end{equation}
where
\begin{eqnarray}
	\overline{S}&=&S+\frac{1}{\nu}\left[(\Gamma-\underline{\mu}_{\Gamma})'\underline{\Omega}_{\Gamma}^{-1}(\Gamma-\underline{\mu}_{\Gamma})+2(A_{\star}-\underline{\mu}_{\star})(A_{\star}-\underline{\mu}_{\star})'+(A_1-\underline{\mu}_1)\underline{\Omega}_1^{-1}(A_1-\underline{\mu}_1)'+\right.\nonumber\\
	&+&\left.(A_2-\underline{\mu}_2)\underline{\Omega}_2^{-1}(A_2-\underline{\mu}_2)'\right]+E'E,
\end{eqnarray}

\item the inverse gamma distribution for $\nu$ (if it is estimated)
\begin{equation}
p(\nu|\cdot,y)=iG\left(\overline{s}_{\nu},\overline{\Omega}_{bI}\right),
\label{eq:iGnu}
\end{equation}
where $\overline{n}_{\nu}=\underline{n}_{\nu}+\frac{n}{2}[n(k-4)+l+r_1+r_2+2r_3]$ and
\begin{eqnarray}
	\overline{s}_{\nu}&=&\underline{s}_{\nu}+\frac{1}{2}tr\left\{\Sigma^{-1}\left[(\Gamma-\underline{\mu}_{\Gamma})'\underline{\Omega}_{\Gamma}^{-1}(\Gamma-\underline{\mu}_{\Gamma})+2(A_{\star}-\underline{\mu}_{\star})(A_{\star}-\underline{\mu}_{\star})'+\right.\right.\nonumber\\
	&+&\left.\left.(A_1-\underline{\mu}_1)\underline{\Omega}_1^{-1}(A_1-\underline{\mu}_1)'+(A_2-\underline{\mu}_2)\underline{\Omega}_2^{-1}(A_2-\underline{\mu}_2)'\right]\right\}.
\end{eqnarray}

\item the matrix normal distribution for $\Gamma$:
\begin{equation}
p(\Gamma|\cdot,y)=f_{mN}\left(\overline{\mu}_{\Gamma},\Sigma,\overline{\Omega}_{\Gamma}\right),
\label{eq:mNG}
\end{equation} 
where $\overline{\Omega}_{\Gamma}=(\frac{1}{\nu}\underline{\Omega}_{\Gamma}^{-1}+Z_4'Z_4)^{-1}$,\\
 $\overline{\mu}_{\Gamma}=\overline{\Omega}_{\Gamma}\left[\frac{1}{\nu}\underline{\Omega}_{\Gamma}^{-1}\underline{\mu}_{\Gamma}+Z_4'\left(Z_0-Z_1B_1A_1'-Z_2B_2A_2'-2Re(Z_3\bar{B}_{\star}A_{\star}')\right)\right]$,

\item the matrix normal distribution for $A_1$:
\begin{equation}
p(A_1|\cdot,y)=f_{mN}\left(\overline{\mu}_1,\overline{\Omega}_1,\Sigma\right),
\label{eq:mNA1}
\end{equation} 
where $\overline{\Omega}_1=(\frac{1}{\nu}\underline{\Omega}_1^{-1}+B_1'Z_1'Z_1B_1)^{-1}$,\\
 $\overline{\mu}_1=\left[\frac{1}{\nu}\underline{\Omega}_1^{-1}\underline{\mu}_1+\left(Z_0-Z_2B_2A_2'-2Re(Z_3\bar{B}_{\star}A_{\star}')-Z_4\Gamma\right)'Z_1B_1\right]\overline{\Omega}_1$,

\item the matrix normal distribution for $A_2$:
\begin{equation}
p(A_2|\cdot,y)=f_{mN}\left(\overline{\mu}_2,\overline{\Omega}_2,\Sigma\right),
\label{eq:mNA2}
\end{equation} 
where $\overline{\Omega}_2=(\frac{1}{\nu}\underline{\Omega}_2^{-1}+B_2'Z_2'Z_2B_2)^{-1}$,\\
 $\overline{\mu}_2=\left[\frac{1}{\nu}\underline{\Omega}_2^{-1}\underline{\mu}_2+\left(Z_0-Z_1B_1A_1'-2Re(Z_3\bar{B}_{\star}A_{\star}')-Z_4\Gamma\right)'Z_2B_2\right]\overline{\Omega}_2$,

\item the matrix normal distribution for $A_{RI}=\left(\begin{array}{c}A_R'\\A_I'\end{array}\right)$:
\begin{equation}
p(A_{RI}|\cdot,y)=f_{mN}\left(\overline{\mu}_{RI},\Sigma,\overline{\Omega}_{RI}\right),
\label{eq:mNARI}
\end{equation} 
where $\overline{\Omega}_{RI}=(\frac{2}{\nu}I_{2r_3}+X_{\star}'X_{\star})^{-1}$,\\
 $\overline{\mu}_{RI}=\overline{\Omega}_{RI}\left[\frac{2}{\nu}\underline{\mu}_{RI}+2X_{\star}'\left(Z_0-Z_1B_1A_1'-Z_2B_2A_2'-Z_4\Gamma\right)\right]$, $\underline{\mu}_{RI}=\left(\begin{array}{c}\underline{\mu}_{\star R}'\\\underline{\mu}_{\star I}'\end{array}\right)$, \\
$X_{\star}=2\left[Re(Z_3\bar{B})-Im(Z_3\bar{B})\right]=2\left[(Z_{31}-Z_{32})B_R-(Z_{31}+Z_{32})B_I\right]$,

\item the normal distribution for the vector $b_1=vec(B_1)$:
\begin{equation}
p(b_1|\cdot,y)=f_N\left(\overline{\mu}_{b1},\overline{\Omega}_{b1}\right),
\label{eq:Nb1}
\end{equation}
where $\overline{\Omega}_{b1}=\left[(m_1I_{r_1}\otimes P_1^{-1})+(A_1'\Sigma^{-1}A_1\otimes Z_1'Z_1)\right]^{-1}$,\\
 $\overline{\mu}_{b1}=\overline{\Omega}_{b1}vec\left[Z_1'\left(Z_0-Z_2B_2A_2'-2Re(Z_3\bar{B}_{\star}A_{\star}')-Z_4\Gamma\right)\Sigma^{-1}A_1\right]$,

\item the normal distribution for the vector $b_2=vec(B_2)$:
\begin{equation}
p(b_2|\cdot,y)=f_N\left(\overline{\mu}_{b2},\overline{\Omega}_{b2}\right),
\label{eq:Nb2}
\end{equation}
where $\overline{\Omega}_{b2}=\left[(m_2I_{r_2}\otimes P_2^{-1})+(A_2'\Sigma^{-1}A_2\otimes Z_2'Z_2)\right]^{-1}$,\\
 $\overline{\mu}_{b2}=\overline{\Omega}_{b2}vec\left[Z_2'\left(Z_0-Z_1B_1A_1'-2Re(Z_3\bar{B}_{\star}A_{\star}')-Z_4\Gamma\right)\Sigma^{-1}A_2\right]$,

\item the normal distribution for the vector $b_R=vec(B_R)$
\begin{equation}
p(b_R|\cdot,y)=f_N\left(\overline{\mu}_{bR},\overline{\Omega}_{bR}\right),
\label{eq:NbR}
\end{equation}
where $\overline{\Omega}_{bR}=\left[x_{bR}'(\Sigma^{-1}\otimes I_T)x_{bR}+2(m_3I_{r3}\otimes P_{\star R}^{-1})\right]^{-1}$,\\
 $\overline{\mu}_{bR}=2\overline{\Omega}_{bR}vec\left[Z_{31}'Y_{bR}\Sigma^{-1}A_I-Z_{32}'Y_{bR}\Sigma^{-1}A_R\right]$,\\
$x_{bR}=2(A_I\otimes Z_{31})-2(A_R\otimes Z_{32})$,\\
$Y_{bR}=Z_0-Z_1B_1A_1'-Z_2B_2A_2'-Z_4\Gamma+2Z_{31}B_IA_R'+2Z_{32}B_IA_I'$,

\item the normal distribution for the vector $b_I=vec(B_I)$
\begin{equation}
p(b_I|\cdot,y)=f_N\left(\overline{\mu}_{bI},\overline{\Omega}_{bI}\right),
\label{eq:NbI}
\end{equation}
where $\overline{\Omega}_{bI}=\left[x_{bI}'(\Sigma^{-1}\otimes I_T)x_{bI}+2(m_3I_{r3}\otimes (P_{\star R}+P_{\star I}P_{\star R}^{-1}P_{\star I})^{-1})\right]^{-1}$,\\
 $\overline{\mu}_{bI}=2\overline{\Omega}_{bI}vec\left[m_3(P_{\star R}+P_{\star I}P_{\star R}^{-1}P_{\star I})^{-1}P_{\star I}P_{\star R}^{-1}B_R-Z_{31}'Y_{bI}\Sigma^{-1}A_R-Z_{32}'Y_{bI}\Sigma^{-1}A_I\right]$,\\
$x_{bI}=-2(A_R\otimes Z_{31})-2(A_I\otimes Z_{32})$,\\
$Y_{bI}=Z_0-Z_1B_1A_1'-Z_2B_2A_2'-Z_4\Gamma-2Z_{31}B_RA_I'+2Z_{32}B_RA_R'$,
\end{itemize}

Having the set of full conditional posterior distributions, the pseudo-random sample from the joint posterior distribution may be obtained with the help of the Gibbs sampler, similarly as \citet{Koop_al2009} in CI(1,1) case.\\
In the first step the initial values are proposed - $\Sigma^{(0)},\ \nu^{(0)},\ \Gamma^{(0)},\ A_1^{(0)},\ B_1^{(0)},\ A_2^{(0)},\ B_2^{(0)},\ A_{\star}^{(0)},\ B_{\star}^{(0)}$, then the following steps are reiterated:
\begin{itemize}
\item draw $\Sigma^{(s)}$ from the inverse Wishart distribution (\ref{eq:iW}),
\item draw $\nu^{(s)}$ from the inverse gamma distribution (\ref{eq:iGnu}) - if it is estimated,
\item draw $\Gamma^{(s)}$ from the matrix normal distribution (\ref{eq:mNG}),
\item draw $A_1^{(s)}$ from the matrix normal distribution (\ref{eq:mNA1}),
\item draw $vec(B_1)^{(s)}$ from the normal distribution (\ref{eq:Nb1}) and reshape it to obtain $B_1$,
\item obtain $\beta_1^{(s)}$ and $\alpha_1^{(s)}$ as $\beta_1^{(s)}=B_1^{(s)}(B_1^{(s)'}B_1^{(s)})^{-\frac{1}{2}}$ and $\alpha_1^{(s)}=A_1^{(s)}(B_1^{(s)'}B_1^{(s)})^{\frac{1}{2}}$,
\item draw $A_2^{(s)}$ from the matrix normal distribution (\ref{eq:mNA2}),
\item draw $vec(B_2)^{(s)}$ from the normal distribution (\ref{eq:Nb2})  and reshape it to obtain $B_2$,
\item obtain $\beta_2^{(s)}$ and $\alpha_2^{(s)}$ as $\beta_2^{(s)}=B_2^{(s)}(B_2^{(s)'}B_2^{(s)})^{-\frac{1}{2}}$ and $\alpha_2^{(s)}=A_2^{(s)}(B_2^{(s)'}B_2^{(s)})^{\frac{1}{2}}$,
\item draw $A_R^{(s)}$ and $A_I^{(s)}$ from the matrix normal distribution (\ref{eq:mNARI}),
\item draw $vec(B_R)^{(s)}$ from the normal distribution (\ref{eq:NbR})  and reshape it to obtain $B_R$,
\item draw $vec(B_I)^{(s)}$ from the normal distribution (\ref{eq:NbI})  and reshape it to obtain $B_I$,
\item set $A_{\star}^{(s)}=A_R+iA_I$ and $B_{\star}^{(s)}=B_R+iB_I$,
\item obtain $\beta_{\star}^{(s)}$ and $\alpha_{\star}^{(s)}$ as $\beta_{\star}^{(s)}=B_{\star}^{(s)}(\bar{B}_{\star}^{(s)'}B_{\star}^{(s)})^{-\frac{1}{2}}$ and $\alpha_{\star}^{(s)}=A_{\star}^{(s)}(\bar{B}_{\star}^{(s)'}B_{\star}^{(s)})^{\frac{1}{2}}$,
\item check the non-explosive condition and if it is fulfilled keep the draws and increase the iteration counter.\\
Note that in models with unit roots at various frequencies the non-explosive condition should be examined carefully by taking into account the explicit number of unit roots at particular frequencies.  
\end{itemize}
The square root of the complex Hermitian matrix $(\bar{B}_{\star}^{(s)'}B_{\star}^{(s)})^{\frac{1}{2}}$, may be obtained with the Newton's method proposed by \citet{Highman1986}.

\section{Point estimation of the cointegration space}
\label{sec:point}

Information on the cointegration spaces obtained from the data may be summarized in the point estimate of these spaces and the measure of their posterior distributions' dispersion. \citet{Villani2006} proposed to employ the Frobenius (Hilbert-Schmidt) matrix norm to build the loss function needed to point estimation of the real cointegration space. The same approach can be used to estimate complex spaces (see e.g. \citealp{Srivastava2000}).\\
Employing the Frobenius matrix norm $\|A\|_F=(tr(\bar{A}'A))^{\frac{1}{2}}$ to the projection matrices we can built the loss function of the following form:
\begin{equation}
l(\beta,\tilde{\beta})=\|\beta\bar{\beta}'-\tilde{\beta}\bar{\tilde{\beta}}'\|_F^2=2(r-tr(\beta\bar{\beta}'\tilde{\beta}\bar{\tilde{\beta}}')),
\label{eq:loss}
\end{equation}
where $r$ denotes the number of cointegrating vectors.\\
The loss function (\ref{eq:loss}) reaches its minimum in:
\begin{equation}
\hat{\beta}=\left(\begin{array}{cccc}\nu_1&\nu_2&\dots&\nu_r\end{array}\right),
\label{eq:point}
\end{equation}
where $\nu_i$ ($i=1,2,\dots,r$) is the eigenvector of the matrix $E(\beta\bar{\beta}')$ corresponding to its $i$th largest eigenvalue (see \citealp{Chikuse2003}, \citealp{Villani2006}).\\
The numerical realization of $\hat{\beta}$ is obtained with the use of the pseudo-random sample from the posterior distribution of $\beta$, $\{\beta^{(s)},\ s=1,2,\dots,S\}$, by approximating $E(\beta\bar{\beta}')$ as $\frac{1}{S}\sum_{s=1}^S\beta^{(s)}\bar{\beta}^{(s)'}$.\\
Following \citet{Villani2006} we use the projective Frobenius span variation:
\begin{equation}
\tau_{sp(\beta)}^2=\frac{r-\sum_{i=1}^r\lambda_i}{r(m-r)/m}\in[0,1],
\label{eq:variation}
\end{equation}
where $\lambda_i$ is the $i$th largest eigenvalue of $E(\beta\bar{\beta}')$. The measure $\tau_{sp(\beta)}^2$ reaches its minimum value when the distribution is degenerated, whereas it hits the maximum value for the uniform distribution over the complex Grassmann manifold ($\mathbb{G}_{r,m-r}$).\\

\section{Simulation experiment}
\label{sec:MCexp}
As a first illustration of the proposed methods, we perform a small simulation study. We use one of the data generating processes proposed by \cite{Cubadda_Omtzigt2005}.
We simulate 250 data points from:
\begin{eqnarray}
\Delta_4y_t&=&A_1B_1'y_t^{(1)}+A_2B_2'y_t^{(2)}+A_{\star}\bar{B}_{\star}'y_t^{(3)}+\bar{A}_{\star}B_{\star}'\bar{y}_t^{(3)}+\nonumber\\
&&{}+\Gamma_1\Delta_4y_{t-1}+\varepsilon_t,\quad \varepsilon_t\sim iiN(0,\Sigma),
\label{eqn:simulation}
\end{eqnarray}
where $B_1=B_2=\left(\begin{array}{r}1\\-1\end{array}\right)$, $B_{\star}=\left(\begin{array}{r}1\\0\end{array}\right)+i\left(\begin{array}{r}0\\1\end{array}\right)$, $A_1=\left(\begin{array}{r}-0.2\\0\end{array}\right)$, $A_2=\left(\begin{array}{r}0.2\\0\end{array}\right)$, $A_{\star}=i\left(\begin{array}{r}0.1\\0\end{array}\right)$, $\Gamma=\left(\begin{array}{rr}0.1&-0.1\\-0.2&0.17\end{array}\right)$ and $\Sigma=\left(\begin{array}{rr}1&-\frac{\sqrt{2}}{4}\\-\frac{\sqrt{2}}{4}&0.5\end{array}\right)$.\\
Initial values are set to zeros, then the first 50 points are discarded, so we are left with 200 modeled data points.\\
We impose the following priors:
\begin{itemize}
	\item $\Sigma\sim iW(0.1I_2,4),$
	\item $\Gamma|\Sigma\sim mN\left(\mathbf{0},\Sigma,\nu I_2\right),$
	\item $A_1|\Sigma\sim mN(\mathbf{0}_{n\times r_1},\nu I_{r_1},\Sigma),$
	\item $B_1\sim mN(0,\frac{1}{m_1}I_{r_1},0.1I_2),$
	\item $A_2|\Sigma\sim mN(\mathbf{0}_{n\times r_2},\nu I_{r_2},\Sigma),$
	\item $B_2\sim mN(0,\frac{1}{m_2}I_{r_2},0.1I_2),$
	\item $A_{\star}|\Sigma\sim mCN(\mathbf{0}_{n\times r_3},\nu I_{r_3},\Sigma),$
	\item $B_{\star}\sim mCN(\mathbf{0}_{m_r\times r_3},I_{r_3},0.1I_2),$
	\item $\nu\sim iG(1,1)$.
\end{itemize}
Table \ref{tab:pMy_sim} gathered the results of the Bayesian model comparison (see appendix for more information about the employed methods).
\begin{table}[H]
	\caption{The most probable models in the simulation experiment ($p(M_{d,s,r_1,r_2,r_3}|y)>0.001$).}
	\label{tab:pMy_sim}
	\begin{center}
		\begin{tabular}{p{1.5 cm}p{1.5 cm}p{1.5 cm}p{1.5 cm}p{1.5 cm}c}
			\hline
			$d$&$s$&$r_1$&$r_2$&$r_3$&$p(M_{d,s,r_1,r_2,r_3}|y)$\\
			\hline
			3&	0&	1&	1&	1&	0.606\\
			4&	0&	1&	1&	1&	0.383\\
			2&	0&	1&	1&	1&	0.008\\
			\hline
		\end{tabular}
	\end{center}
{\small\textit{Note:}  The symbol $d$ denotes the type of deterministic components ($d=1$ - a trend restricted to the cointegration space at the zero frequency and an unrestricted constant, $d=2$ - an unrestricted constant, $d=3$ - a constant restricted to the cointegration relation at the zero frequency, $d=4$ - no deterministic components), $s$ can be 0 (no seasonal dummies) or 1 (a model with seasonal dummies), $r_1,r_2,r_3\in\{0,1,2\}$  The true model is $M_{4,0,1,1,1}$ and the prior probability for each of the compared specification is $p(M_{d,s,r_1,r_2,r_3})=\frac{1}{136}\approx 0.007$.\\
\textit{Source:} Own calculations based on 1.5 million draws from the prior distribution.}
\end{table}

There are 3 models with the posterior probability higher than 0.001 and they gathered 0.997 of probability mass. The true model is on the second place with the posterior probability equal 0.383. All the models displayed in the Table \ref{tab:pFy_sim} have proper numbers of cointegrating vectors, have no seasonal dummies and differ only in the type of deterministic components.\\
In Table \ref{tab:pFy_sim} we present the marginal posterior probabilities of models' features.
\begin{table}[H]
	\caption{Posterior probabilities of the model features.}
	\label{tab:pFy_sim}
	\begin{center}
		\begin{tabular}{ccccc}
			\hline\hline
			$p(d=1|y)$&$p(d=2|y)$&$p(d=3|y)$&$p(d=4|y)$&\\
			%\hline
			0.000		&	0.010	&		0.606& 0.384 &\\
			\hline\hline
			$p(s=0|y)$&$p(s=1|y)$&&&\\
			%\hline
			0.999 & 0.001 &&&\\
			\hline\hline
			$p(r_1=0|y)$&$p(r_1=1|y)$&$p(r_1=2|y)$&&\\
			%\hline
			0.000 & 0.998 & 0.002 &&\\
			\hline\hline
			$p(r_2=0|y)$&$p(r_2=1|y)$&$p(r_2=2|y)$&&\\
			%\hline
			0.000 & 1.000 & 0.000 &&\\
			\hline\hline
			$p(r_3=0|y)$&$p(r_3=1|y)$&$p(r_3=2|y)$&&\\
			%\hline
			0.000 & 1.000 & 0.000 &&\\
			\hline\hline			
		\end{tabular}
	\end{center}
{\small\textit{Source:} Own calculations based on 1.5 million draws from the prior distribution.}
\end{table}

The results of models comparison generally correctly point to proper model's characteristics with the exception of the type of deterministic components because the models with the constant restricted to the cointegrating space at zero frequency gathered 0.606 of the posterior probability which is approximately 1.5 more than the whole posterior probability of models without a constant, i.e. the true specification.
In the next step of this small simulation experiment we estimate the true model and present the distances between obtained cointegrating spaces and the assumed ones (Table \ref{tab:sim}).
\begin{table}[H]
	\caption{The results of cointegrated spaces' estimation in the simulation experiment.}
	\label{tab:sim}
	\begin{center}
		\begin{tabular}{ccccc}
			\hline\hline
			frequency&assumed vector ($\beta_i$) &point estimate ($\hat{\beta}_i$) &$\tau^2_{sp(\beta_i)}$&$l^{\frac{1}{2}}(\beta_i,\hat{\beta}_i)$\\
			\hline\hline
			0 ($i=1$)		&	$\left(\begin{array}{c}\frac{\sqrt{2}}{2}\\-\frac{\sqrt{2}}{2}\end{array}\right)$	&	$\left(\begin{array}{c}-0.691\\0.723\end{array}\right)$& 0.0004 &0.032\\
			\hline
			$\pi$ ($i=2$)	&	$\left(\begin{array}{c}\frac{\sqrt{2}}{2}\\-\frac{\sqrt{2}}{2}\end{array}\right)$	&	$\left(\begin{array}{c}-0.715\\0.699\end{array}\right)$& 0.0010 &0.016\\
			\hline
			$\frac{\pi}{2}, \frac{3\pi}{2}$	($i=\star$)	&	$\left(\begin{array}{c}\frac{\sqrt{2}}{2}\\0\end{array}\right)+i\left(\begin{array}{c}0\\\frac{\sqrt{2}}{2}\end{array}\right)$	& 
			$\left(\begin{array}{c}0.029\\0.711\end{array}\right)+i\left(\begin{array}{c}-0.703\\0\end{array}\right)$ & 0.0006 &0.030\\
			\hline\hline			
		\end{tabular}
	\end{center}
{\small \textit{Note:} $\frac{\sqrt{2}}{2}\approx 0.7071$\\
\textit{Source:} Own calculations based on 200000 accepted draws, preceded by 100000 discarded draws form the posterior distribution.}
\end{table}
At the first glance one may notice significant differences between point estimate of the cointegration vectors at the annual frequency and the assumed one, but it should be remembered that the data contain information only about the cointegrating spaces, not the vectors, and the distance between the true and estimated space is very low (see the last column of Table \ref{tab:sim}). Generally, the results of the performed simulation experiment proofed that the proposed procedures works well, as the differences between true and estimated spaces at all frequencies are negligible. Moreover, values of the measure $\tau_{sp(\beta)}^2$ are close to zero, so the posterior distributions of the cointegrating spaces are almost degenerated, which can be expected during the analysis of artificial data. Now we proceed to the employment of the proposed model in the real data analysis.

\section{Empirical illustration}
\label{sec:ill}
\begin{figure}[H]
	\includegraphics[scale=0.5]{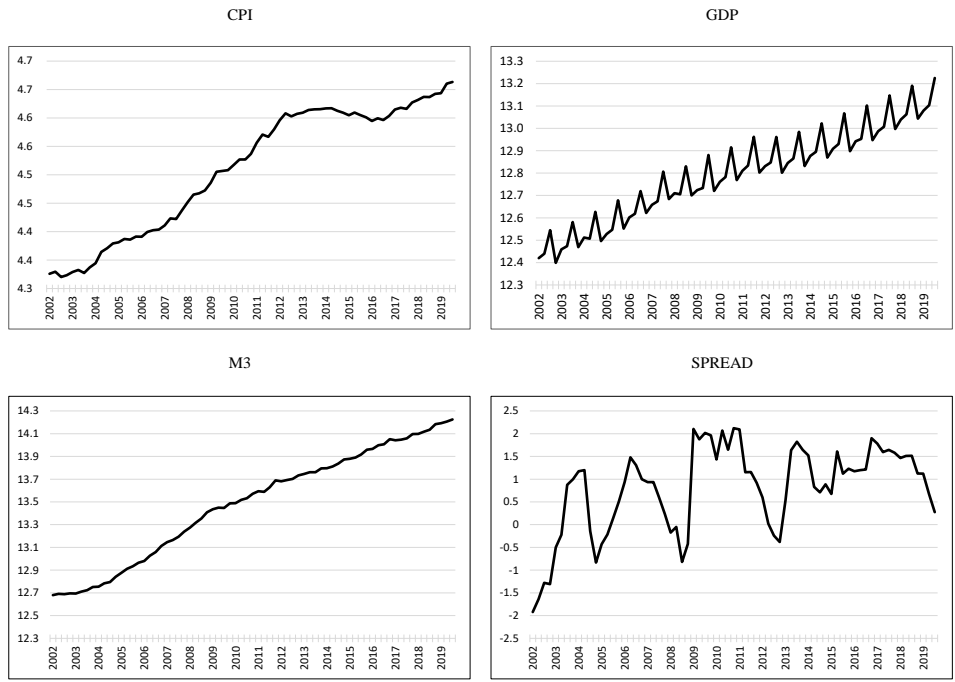}
	\caption{The analyzed data.}
	\label{fig:data}
\end{figure}

\begin{figure}[H]
	\includegraphics[scale=0.5]{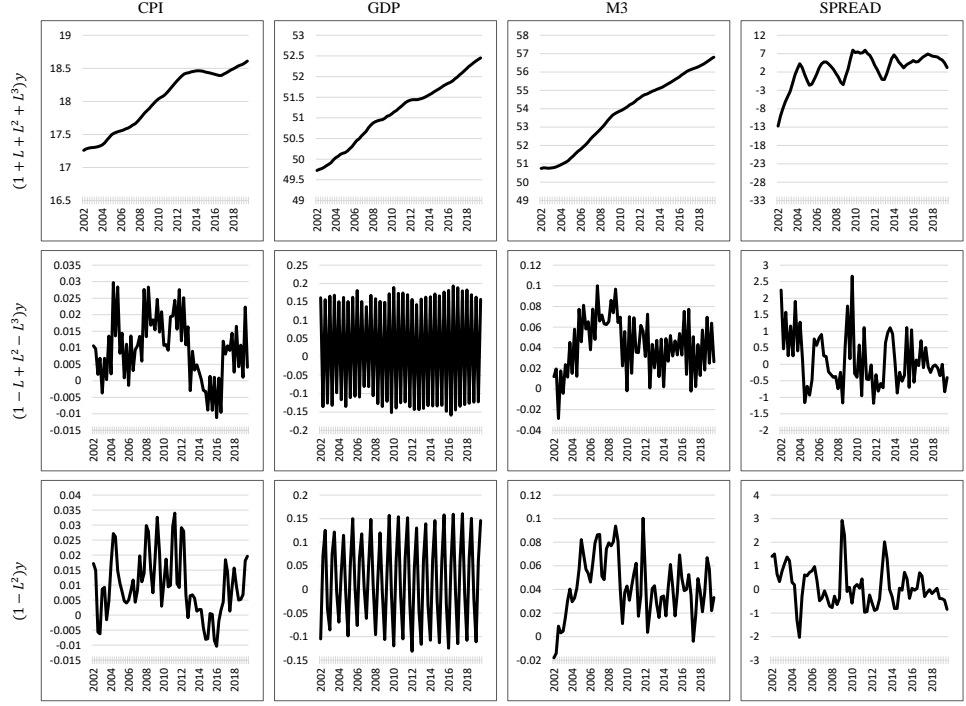}
	\caption{The unit root transformations od the analyzed data.}
	\label{fig:data_trans}
\end{figure}

In the empirical analysis, we will consider the four-dimensional time series consisted of GDP in constant prices from 2010, consumer price index ($2015=100$), the broad monetary aggregate M3, and the spread between long- and short-term interest rates approximated as the difference between 10-Year Bond Yield and 3-months WIBOR. The quarterly data cover the period 2002Q1 - 2019Q4.\\
A similar model was analyzed by \cite{Kotlowski2005}.\\
Figure \ref{fig:data} depicts the analyzed time series. The seasonality may be observed in the paths of all the considered time series, whereas the strongest seasonal pattern is visible in GDP. The trending behavior of the series is also noticeable.
Following \cite{Franses1994} and \cite{Granger_al1993} we present also the unit transformation of the analyzed time series (Figure \ref{fig:data_trans}), where the trending behavior and seasonal patterns are more visible. The seasonal variations of GDP differ from the changes observed at the rest of the series. 
To fully define the Bayesian seasonally cointegrated VAR model we impose the following priors:
\begin{itemize}
	\item $\Sigma\sim iW(0.1I_4,6),$
	\item $\Gamma|\Sigma\sim mN\left(\mathbf{0},\Sigma,\nu I_{4+l}\right),$ where $l$ denotes the number of dummies outside cointegration spaces,
	\item $A_1|\Sigma\sim mN(\mathbf{0}_{n\times r_1},\nu I_{r_1},\Sigma),$
	\item $B_1\sim mN(0,\frac{1}{m_1}I_{r_1},0.1I_4),$
	\item $A_2|\Sigma\sim mN(\mathbf{0}_{n\times r_2},\nu I_{r_2},\Sigma),$
	\item $B_2\sim mN(0,\frac{1}{m_2}I_{r_2},0.1I_4),$
	\item $A_{\star}|\Sigma\sim mCN(\mathbf{0}_{n\times r_3},\nu I_{r_3},\Sigma),$
	\item $B_{\star}\sim mCN(\mathbf{0}_{m_r\times r_3},I_{r_3},0.1I_4),$
	\item $\nu\sim iG(1,1)$.
\end{itemize}
In order to check the nature of the analyzed series we start the analysis by the Bayesian model comparison. The models may differ in the number of cointegrating relations ($r_j\in\{0,1,2,3,4\}$ for $j=1,2,3$) at zero ($j=1$), $\pi$ ($j=2$) and $\frac{\pi}{2}$, $\frac{3\pi}{2}$ ($j=3$) frequencies. 
We consider models with a linear trend restricted to the cointegration space at zero frequency and with an unrestricted constant ($d=1$), models with an unrestricted constant ($d=2$), specifications with a constant restricted to the cointegration space at zero frequency ($d=3$), and models without constant ($d=4$). The models without ($s=0$) and with ($s=1$) seasonal dummies were examined. Each of the considered specifications has five lags in VAR representation. After excluding non-possible feature combinations and leaving in the set one representation of the observationally equivalent models we are left with 784 pairwise different models. We assume equal prior probability of each specification, i.e. $p(M_{d,s,r_1,r_2,r_3})=\frac{1}{784}\approx 0.0013$. Note that imposing uniform probability on the models' space does not lead to uniform prior distribution for the model features (see the numbers in parentheses in Table \ref{tab:pFy}).\\
The models with posterior probability higher than 0.01 are displayed in Table \ref{tab:pMy}. Almost all of the listed models assume that the analyzed times series may be treated as a realization of the process with 3 or 4 bi-annual relations (note that 4 means stability at $\pi$ frequency). The models ranked at the first and second place assume one long-run relationship and an unrestricted constant. They differ only in the number of cointegrating vectors at the bi-annual frequency (4 or 3).  Note that according to the results displayed in Table \ref{tab:pFy}, models with two relations at 0 frequency are the most probable and gathered 0.409 of the posterior probability, whereas models with one relation 0.363. Models assuming stability at bi-annual frequency together obtained 0.432 of the probability mass, so there is still evidence of cointegration at this frequency as the whole posterior probability of models assuming it is higher and equals 0.562. The posterior probability of the number of cointegrating relations at annual frequency is also diffused, but the whole probability of cointegration equals 0.949, so there is a strong confirmation of the existence of cointegration of this type.
\begin{table}[H]
	\caption{The most probable models ($p(M_{d,s,r_1,r_2,r_3}|y)>0.01$).}
	\label{tab:pMy}
	\begin{center}
		\begin{tabular}{p{1.5 cm}p{1.5 cm}p{1.5 cm}p{1.5 cm}p{1.5 cm}c}
			\hline
			$d$&$s$&$r_1$&$r_2$&$r_3$&$p(M_{d,s,r_1,r_2,r_3}|y)$\\
			\hline
	2&	0&	1&	4&	3&	0.053\\
	2&	0&	1&	3&	3&	0.048\\
	2&	0&	0&	4&	3&	0.032\\
	4&	0&	2&	4&	3&	0.031\\
	3&	0&	2&	4&	3&	0.031\\
	2&	0&	1&	3&	2&	0.031\\
	2&	0&	1&	2&	3&	0.030\\
	2&	0&	2&	4&	3&	0.029\\
	2&	0&	1&	4&	2&	0.027\\
	2&	0&	0&	3&	3&	0.023\\
	2&	0&	1&	2&	2&	0.021\\
	3&	0&	2&	3&	3&	0.020\\
	4&	0&	2&	3&	3&	0.020\\
	2&	0&	0&	4&	2&	0.016\\
	3&	0&	2&	4&	2&	0.015\\
	2&	0&	0&	3&	2&	0.015\\
	2&	0&	2&	3&	3&	0.014\\
	2&	0&	1&	1&	3&	0.014\\
	2&	0&	0&	2&	3&	0.014\\
	3&	0&	2&	2&	3&	0.014\\
	4&	0&	2&	4&	2&	0.014\\
	4&	0&	2&	2&	3&	0.014\\
	2&	0&	1&	3&	1&	0.013\\
	2&	0&	2&	4&	2&	0.013\\
	3&	0&	2&	3&	2&	0.012\\
	4&	0&	2&	3&	2&	0.012\\
	2&	0&	1&	1&	2&	0.011\\
			\hline
		\end{tabular}
	\end{center}
{\small \textit{Source:} Own calculations based on 1.5 million draws from the prior distribution.}
\end{table}

\begin{table}[H]
	\caption{Posterior probabilities of models' features.}
	\label{tab:pFy}
	\begin{center}
		\begin{tabular}{ccccc}
			\hline\hline
			$p(d=1|y)$&$p(d=2|y)$&$p(d=3|y)$&$p(d=4|y)$&\\
			%\hline
			0.053		&	0.570	&		0.196& 0.181 &\\
			(0.188)		&	(0.312)	&		(0.188)		&	(0.312) &\\
			\hline\hline
			$p(s=0|y)$&$p(s=1|y)$&&&\\
			%\hline
			0.965 & 0.035 &&&\\
			(0.510) & (0.490) &&&\\
			\hline\hline
			$p(r_1=0|y)$&$p(r_1=1|y)$&$p(r_1=2|y)$&$p(r_1=3|y)$&$p(r_1=4|y)$\\
			%\hline
			0.143 & 0.363 & 0.409 & 0.085 & 0.000\\
			(0.125) & (0.250) & (0.250) & (0.250) & (0.125)\\
			\hline\hline
			$p(r_2=0|y)$&$p(r_2=1|y)$&$p(r_2=2|y)$&$p(r_2=3|y)$&$p(r_2=4|y)$\\
			%\hline
			0.006 & 0.090 & 0.189 & 0.283 & 0.432\\
			(0.184) & (0.204) & (0.204) & (0.204) & (0.204)\\
			\hline\hline
			$p(r_3=0|y)$&$p(r_3=1|y)$&$p(r_3=2|y)$&$p(r_3=3|y)$&$p(r_3=4|y)$\\
			%\hline
			0.000 & 0.125 & 0.305 & 0.519 & 0.051\\
			(0.184) & (0.204) & (0.204) & (0.204) & (0.204)\\
			\hline\hline			
		\end{tabular}
	\end{center}
{\small \textit{Note:} In parentheses - the assumed prior probabilities.\\
\textit{Source:} Own calculations based on 1.5 million draws from the prior distribution.}
\end{table}

\begin{figure}[H]
	\includegraphics[scale=0.5]{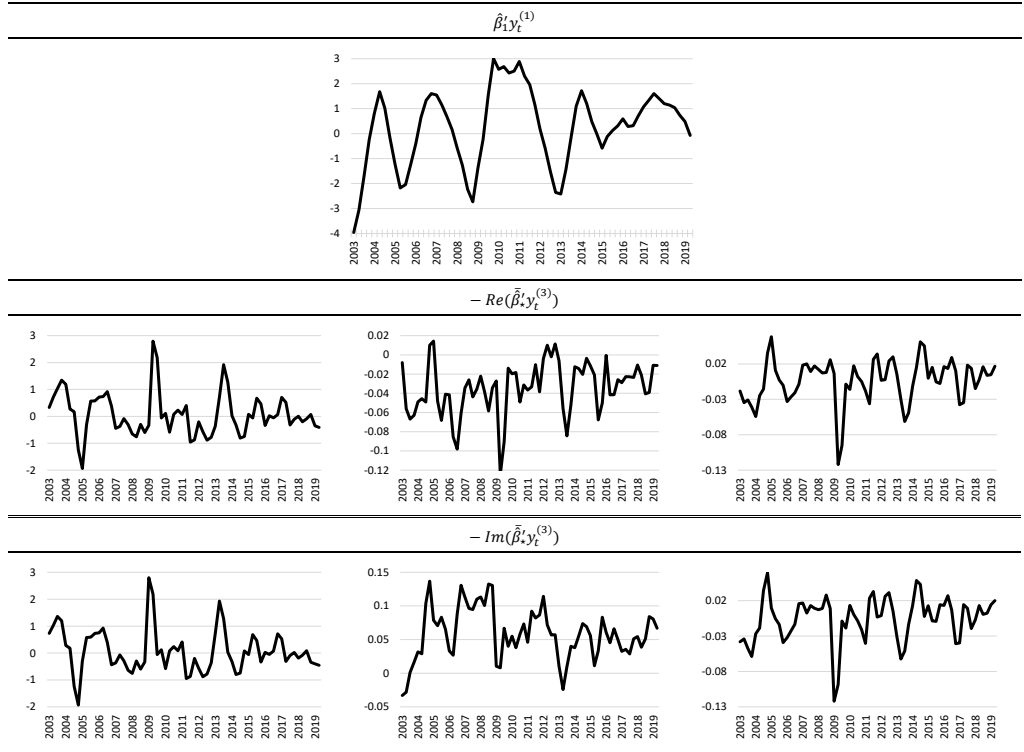}
	{\small \textit{Source:} Own calculations.}
	\caption{Deviations from the obtained cointegrating relations.}
	\label{fig:relations}
\end{figure}

\begin{figure}[H]
	\includegraphics[scale=0.5]{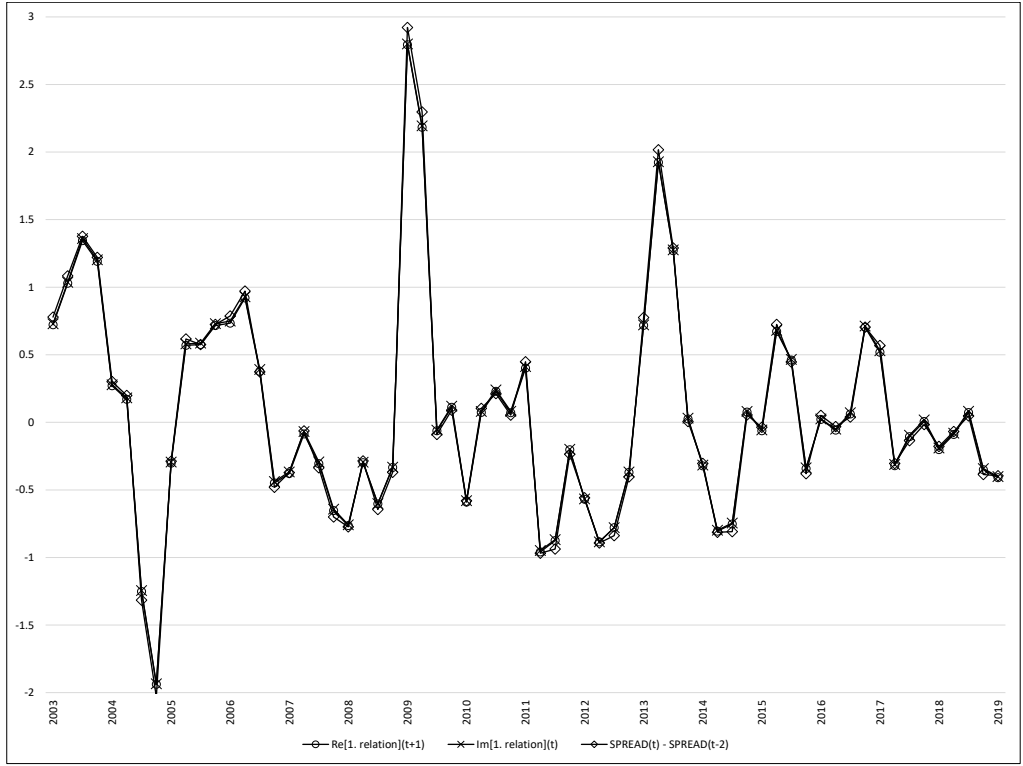}
	{\small \textit{Source:} Own calculations.}
	\caption{Comparison of $SPREAD_t-SPREAD_{t-2}$ and deviations from the $1^{st}$ cointegrating relation at the annual frequency.}
	\label{fig:comp}
\end{figure}

The discussion of nature of the analyzed time series will be complemented by point estimation of the cointegration spaces in the most probable model - $M_{2,0,1,4,3}$.\\
\begin{equation}
\hat{\beta}_1=\left(\begin{array}{r}-0.040\\0.551\\-0.546\\0.630\end{array}\right), \tau^2_{sp(\beta_1)}=0.536
\end{equation}
The posterior distribution of the cointegration space at zero frequency is quite diffuse, as the measure $\tau^2_{sp(\beta_1)}$ is roughly in the middle of the interval $[0, 1]$. The visual inspection of deviations from the obtained cointegrating relation (graphed at Figure \ref{fig:relations}) confirms its stationarity.\\
The estimation's results of the cointegration space at the annual frequency are a bit surprising. The measure $\tau^2_{sp(\beta_{\star})}$ is just over 0.9, so the posterior distribution of the cointegration space is almost uniform.
\begin{eqnarray}
&&\hat{\beta}_{\star}=\left(\begin{array}{rrr}-0.053&0.055&-0.980\\0.140&0.083&0.058\\0.007&0.280&0.147\\0.964&-0.032&-0.038\end{array}\right)+
i\left(\begin{array}{rrr}-0.027&-0.151&-0.007\\-0.217&-0.126&0.112\\0.012&-0.934&-0.021\\0&0&0\end{array}\right),\nonumber\\
&&\tau^2_{sp(\beta_1)}=0.930
\end{eqnarray}
Deviations from the obtained relations seem to be stationary, but by a closer look at the first relation we can notice that its real and imaginary parts are almost the same \footnote{Similar remark applies to the $3^{rd}$ relation}. Moreover, these paths resemble the dynamics of transformations for SPREAD, i.e. $SPREAD_t-SPREAD_{t-2}$ (see Figure \ref{fig:comp}). Such results may indicate for stationarity of SPREAD at the annual frequency. This hypothesis may be checked by the Bayesian model comparison between models with such stationarity restriction imposed and without it, but this is left for further research.
\section{Conclusions}
\label{sec:conc}
In this paper, the Bayesian seasonally cointegrated vector error correction model for quarterly data was introduced. The empirical usefulness of the discussed methods was illustrated by the analysis of four-dimensional time series. Results of the model comparison indicate that data support the hypothesis of cointegration at zero and annual frequency. There is also evidence of cointegration at bi-annual frequency.\\
This paper focuses only on Bayesian model comparison and point estimation in the most probable model, but as the posterior distribution of the model's specification is diffused it would be useful to take advantage of the Bayesian knowledge pooling in the set of the most probable models. Moreover, as there are papers demonstrating risk in omitting seasonal cointegration (see Introduction), the presented research can also be extended by comparison of forecasts performed in the set of models taking into account relationships at seasonal frequencies and those allowing for only long-run relations. A similar comparison may be performed for structural analyses. Such examination is left for further research.

\section*{Appendix - the Bayesian model comparison}
 Bayesian methodology enables to compare different model specifications through the posterior probability (see e.g. \citealp{Zellner1971}, \citealp{Osiewalski2001}, \citealp{Pajor2017}). The model with the highest posterior probability is usually considered the best, but having posterior probabilities of the considered models, further analyses may be based not only on that one chosen specifications but with the use of Bayesian knowledge pooling within the group of models with the highest posterior probabilities.\\
The seasonally cointegrated VAR models may differ in the lag length ($k$), the type of the deterministic trends ($d$) and seasonal dummies ($s$), and  in the number of the cointegrating relations ($r_1$, $r_2$, $r_3$). \\
Using the Bayes theorem, we can evaluate the posterior probability of each model from the set of non-nested competing Bayesian models $\{M_{\xi}: {\xi} = (k, d, s, r_1, r_2, r_3) \in\Xi\}$, where:
\begin{equation}
M_{\xi}=p_{\xi}(y,\theta_{(\xi)})=p_{\xi}(\theta_{(\xi)})p_{\xi}(y|\theta_{(\xi)}),\quad \xi\in\Xi,
\end{equation}
$y$ denotes the data, $\theta_{(\xi)}\in\Theta_{(\xi)}$ is the vector of the parameters of model $M_{\xi}$  and $p_{\xi}(\theta_{(\xi)})$ is the prior density.\\
The posterior probability of model $M_{\xi}$ is:
\begin{equation}
p(M_{\xi}|y) = \frac{p(M_{\xi})p(y|M_{\xi})}{\sum_{\zeta\in\Xi}p(M_{\zeta})p(x|M_{\zeta})},
\label{eq:pY}
\end{equation}
where
\begin{equation}
p(y|M_{\xi}) = \int_{\Theta_{(\xi)}}p_{\xi}(y|\theta_{(\xi)})p_{\xi}(\theta_{(\xi)})\ud\theta_{(\xi)},\quad \xi\in\Xi
\label{eq:MDD}
\end{equation}
is the marginal data density under model $M_{\xi}$ and $p(M_{\xi})$ is the prior probability of the model $M_{\xi}$, so to obtain the posterior probabilities of the compared models (\ref{eq:pY}) the researcher has to evaluate the integral (\ref{eq:MDD}). In most cases it is impossible to obtain it analytically, so one has to use the numerical method. In this paper, we integrate analytically the covariance matrix $\Sigma$, the matrix $\Gamma$, and the matrices $A$s and then the matrices containing the cointegrating vectors will be integrated numerically with the help of the arithmetic mean.\\
Let us write the seasonally cointegrated VAR model in more compact form:
\begin{eqnarray}
Z_0&=&\left(\begin{matrix}Z_1\beta_1&Z_2\beta_2&-2Z_{32}\beta_R-2Z_{31}\beta_I&2Z_{31}\beta_R-2Z_{32}\beta_I&Z_4\end{matrix}\right)\left(\begin{matrix}\alpha_1'\\\alpha_2'\\\alpha_R'\\\alpha_I'\\\Gamma\end{matrix}\right)+E\nonumber\\
&\equiv&\left(\begin{matrix}Z_1B_1&Z_2B_2&-2Z_{32}B_R-2Z_{31}B_I&2Z_{31}B_R-2Z_{32}B_I&Z_4\end{matrix}\right)\left(\begin{matrix}A_1'\\A_2'\\A_R'\\A_I'\\\Gamma\end{matrix}\right)+E\nonumber\\
&=&\tilde{Z}\tilde{\Gamma}+E.
\label{eqn:basic_real}
\end{eqnarray}
where $\tilde{Z}=\left(\begin{matrix}Z_1B_1&Z_2B_2&-2Z_{32}B_R-2Z_{31}B_I&2Z_{31}B_R-2Z_{32}B_I&Z_4\end{matrix}\right)$,\\
$\tilde{\Gamma}=\left(\begin{matrix}A_1&A_2&A_R&A_I&\Gamma'\end{matrix}\right)'$.\\
Using representation (\ref{eqn:basic_real}) and the imposed prior distributions we obtain:
\begin{eqnarray}
p(y|B_1,B_2,B_R,B_I,\nu)&=&\pi^{-\frac{nT}{2}}|\underline{\Omega}_{\tilde{\Gamma}}|^{-\frac{n}{2}}|S|^{\frac{q}{2}}\prod_{i=1}^n\frac{\Gamma[(q+T+1-i)/2]}{\Gamma[(q+1-i)/2]}\times\nonumber\\
&\times&\nu^{-\frac{n}{2}[n(k-4)+l+r_1+r_2+2r_3]}\times\\
&\times&|\overline{\Omega}_{\tilde{\Gamma}}|^{\frac{n}{2}}|S+Z_0'M_{\tilde{Z}}Z_0+R|^{-\frac{1}{2}(q+T)},\nonumber
\label{eqn:pYB}
\end{eqnarray}
where $M_{\tilde{Z}}=I_T-\tilde{Z}(\tilde{Z}'\tilde{Z})^{-1}\tilde{Z}'$, $\overline{\Omega}_{\tilde{\Gamma}}=\left(\frac{1}{\nu}\underline{\Omega}_{\tilde{\Gamma}}^{-1}+\tilde{Z}'\tilde{Z}\right)^{-1}$, $\hat{\tilde{\Gamma}}=(\tilde{Z}'\tilde{Z})^{-1}\tilde{Z}'Z_0$, $R=\frac{1}{\nu}(\underline{\mu}_{\tilde{\Gamma}}-\hat{\tilde{\Gamma}})'\underline{\Omega}_{\tilde{\Gamma}}^{-1}\overline{\Omega}_{\tilde{\Gamma}}\tilde{Z}'\tilde{Z}(\underline{\mu}_{\tilde{\Gamma}}-\hat{\tilde{\Gamma}})$, $\underline{\mu}_{\tilde{\Gamma}}=\left(\begin{array}{ccccc}\underline{\mu}_1&\underline{\mu}_2&\underline{\mu}_{\star R}&\underline{\mu}_{\star I}&\underline{\mu}_{\Gamma}'\end{array}\right)'$, $\underline{\Omega}_{\tilde{\Gamma}}=diag\left(\underline{\Omega}_1, \underline{\Omega}_2, \frac{1}{2}I_{r_3}, \frac{1}{2}I_{r_3}, \underline{\Omega}_{\Gamma}\right)$.\\
To get the marginal data density we have to integrate $\nu$ and the $B_{-}$ matrices numerically. Additionally, we have to remember that the joint prior distribution is truncated by the non-explosive condition, which cuts off a part of the parameters' space. We may calculate $p(y)$ in the non-truncated space and then adjust the normalizing constant of the prior density numerically as the inverse of the fraction of draws lying in this truncated parameter space, $\frac{M}{N}$, where $M$ denotes all draws and $N$ is the number of draws fulfilling the non-explosive condition (see e.g. \citealp{OsiewalskiPipien1999}, \citealp{Osiewalski2001}).

\end{document}